\documentclass[twocolumn,eqsecnum,floatfix,aps,prc]{revtex4}
\usepackage{epsfig}
\usepackage{times}
\usepackage{color}
\def\rhovec{\mbox{\boldmath $\rho$}}

\def\fun#1#2{\lower3.6pt\vbox{\baselineskip0pt\lineskip.9pt
  \ialign{$\mathsurround=0pt#1\hfil##\hfil$\crcr#2\crcr\sim\crcr}}}

\begin{document}


\title {On the possibility of generating a 4-neutron  resonance 
with a {\boldmath $T=3/2$} isospin 3-neutron force}

\author{E. Hiyama}
\affiliation{Nishina Center for Accelerator-Based Science, RIKEN, 
Wako,  351-0198, Japan}

\author{R. Lazauskas}
\affiliation{IPHC, IN2P3-CNRS/Universite Louis Pasteur BP 28, 
F-67037 Strasbourg Cedex 2, France}

\author{J. Carbonell}
\affiliation{Institut de Physique Nucl\'eaire, Universit\'e Paris-Sud, 
IN2P3-CNRS, 91406 Orsay Cedex, France}
\author{M. Kamimura}
\affiliation{Department of Physics, Kyushu University, 
Fukuoka 812-8581, Japan and \\
Nishina Center for Accelerator-Based Science, RIKEN, Wako 351-0198, Japan}

\date{\today}

\begin{abstract}
We consider the theoretical possibility to generate a narrow resonance
in the four neutron system as suggested by a recent experimental result.
To that end,  a phenomenological  $T=3/2$ three neutron  force  is introduced, in addition to a realistic $NN$ interaction. 
We inquire what should be  the strength of the $3n$ force in order to generate such a resonance. 
The reliability  of the three-neutron force in the  $T=3/2$ channel is exmined, by analyzing
its consistency with the low-lying $T=1$ states of $^4$H, $^4$He and $^4$Li and the  $^3{\rm H} + n$ scattering.
 The {\it ab initio} solution of the $4n$ Schr\"{o}dinger equation is obtained
using the complex scaling method   with boundary conditions appropiate to the four-body resonances.
We find that in order to generate narrow $4n$ resonant states  a remarkably  attractive $3N$ force in the $T=3/2$ channel is required.\end{abstract}
\maketitle

\section{INTRODUCTION}

The possibility of detecting a four-neutron (4$n$) structure of any kind -- bound or resonant state --
has intrigued the nuclear physics community for the last fifty years (see
Refs.~\cite{Oglobin_1989,Tilley:1992zz,MMM_JC_2002} for historical reviews).

First of all, many experimental trials have been undertaken to
seek for four-neutron systems, in particular some recurrent claims
of such an observation have been
published~\cite{Marques:2001wh,Marques:2005vz,Chulkov_NPA750_2005,Kisamori_PhD,Kisamori_PRL}
but none of them has really been confirmed
\cite{Didier_NPA805_2008}.


Theoretical studies were mostly concentrated in exploring the
possible existence of  bound multineutrons, agreeing in unison
about the impossibility to observe a bound 4$n$ state
\cite{Bevelacqua_80,BBKE_SJNP41_85,G_YF50_89,SRV_JPG_97,Timofeyuk:2003ya,Pieper:2003dc,Lazauskas_PhD_2003,Lazauskas-3n,Lazauskas-4n}.
None of the known nucleon-nucleon ($NN$) interactions accommodate
such a structure and the required modifications to ensure the 4$n$
binding -- would they be at the 2$N$ or 3$N$ level -- are so
sizeable that  the entire nuclear chart would be strongly perturbed.
Several studies \cite{Timofeyuk:2003ya,Lazauskas_PhD_2003,Pieper:2003dc} indicate that the required additional extrabinding amounts to several
    tens of MeV, much beyond the uncertainties of current nuclear interaction models, including the $3N$ forces

\vskip 0.1cm
A different situation occurs for the 4$n$ resonant states. On the
one hand,  calculation of  resonant states turns out to be a much
more difficult task, both formally and technically. On the other
hand, it is far from trivial to relate the calculated resonant
state parameters, usually their $S$-matrix pole positions, to the
experimental observables. Unless a resonant state is very narrow
and, therefore, a  Breit-Wigner
 parametrization is valid, the experimental
observables have no  straightforward relation with the S-matrix
pole position of the resonance.
In this case, a careful analysis of the reaction mechanism is necessary
in order to establish a relation with the experimental observables.
If a resonance is
found very far from the real energy axis (physical domain)
it will have no significant impact on a physical process.

\vskip 0.1cm
Following this line of reasoning, some calculations based on
semi-realistic $NN$ forces indicated null~\cite{SRV_JPG_97}  or
unlikely evidence~\cite{Timofeyuk:2003ya} for an observable 4$n$
resonance. On the other hand a Green's function Monte Carlo (GFMC)
calculation~\cite{Pieper:2003dc} using the realistic
AV18~\cite{AV18_1995} $NN$ potential and Ilinois-IL2 $3N$ forces
\cite{Pieper2001} suggested a possible (broad) resonance at $E_R$=2 MeV.
This result was  obtained by a  linear extrapolation of tetraneutrons  binding energies,
artificially bound in an external Woods-Saxon potential with $V_0$ depth,  
in the limit $V_0\to0$.
One should note, however,  that in order to determine the position
of a broad resonance by such a procedure, a special functional 
form must be used~\cite{Kukulin},
which  critically depends on the near threshold input values, 
an energy  region where GFMC calculations are difficult to converge.

\vskip 0.1cm
The {\it ab initio} solutions for the 3$n$ and 4$n$ states in the
continuum  were first presented in
Refs.~\cite{Lazauskas-3n,Lazauskas-4n} by solving the
corresponding Faddeev-Yakubovsky (FY) equations~\cite{FY-eq} in
the complex energy plane. The $nn$-interaction used was the charge
dependent  Reid-93 potential from the Nijmegen
group~\cite{REID93_PRC49_1994}. In order to locate the resonance
position of the physical tetraneutron an {\it ad~hoc} 4$n$
interaction was first added to the $nn$  forces with the aim of
artificially creating a 4$n$ bound state. The strength of this
4$n$ term was adiabatically decreased and the trajectory of the
bound state singularity was traced  in the complex energy plane
from the negative real axis to the resonance position  until it
reached its physical point. The conclusion of this work was clear:
none of the examined 4$n$ states ($J^{\pi}=0^{\pm}, 1^{\pm},
2^{\pm}$) could manifest themselves as a near-threshold resonance.
The corresponding pole trajectories moved too far from the real
axis, reached the third energy quadrant and generated large widths
$\Gamma\!\sim\!$ \mbox{ 15 MeV.}

\vskip 0.1cm
A recent experiment on the $^4{\rm He}(^8{\rm He},\mbox{$^8{\rm Be}$})4n$ 
reaction generated an excess of $4n$ events with low energy in the final state.
This observation has been associated with a possible 4$n$ resonance with an estimated energy $E_R=0.83\pm0.65\pm1.25$ MeV
above the $4n$ breakup threshold and an upper limit of width $\Gamma=2.6$ MeV~\cite{Kisamori_PhD,Kisamori_PRL}.
Low statistics, however,  have not allowed one to extract the spin or parity
for  the corresponding state.
It is worth noting that a further analysis of the experimental results of  Ref.~\cite{Marques:2001wh}
concluded that the observed (very few) events  were also compatible with a $E_R=0-2$ MeV tetraneutron resonance  \cite{Marques:2005vz}.

\vskip 0.1cm
In view of the obvious tension  between the theoretical predictions and the last experimental results,
we believe that it would be of  some interest  to reconsider this problem from a different point of view.
In the dynamics of Ref.~\cite{Lazauskas-4n},
 the 3$n$ forces were not included,
 and the  4$n$ force added to the $nn$ potential  was a pure artifact
for binding the system and controlled the singularity when moving
into the resonance region in the complex energy plane.

The $NN$ interaction models are almost perfect in the sense of
$\chi^2/{\rm data}\approx 1$. This was already the case  with the
Nijmegen~\cite{REID93_PRC49_1994}, Bonn~\cite{Bonn} and
Argonne~\cite{AV18_1995} charge symmetry breaking versions,
despite some arbitrariness in their meson contents and couplings.
The remarkable progress that the effective field theory (EFT)
approach to nuclear physics offers, even though it does not
dramatically improve the agreement between the $NN$ models and
$NN$ data, does allow  a high degree of consistency and refinement
that leaves little room for improvement.

The $nn$ interaction part is the least constrained due to the
absence of the experimental data on $nn$ scattering. Still, as
 observed in Ref.~\cite{Lazauskas_PhD_2003,Lazauskas-3n}, the
margin of uncertainty to make the tetraneutron bound or visibly resonant
remains quite limited.

The two-neutron system is virtual state in the
$^1S_0$ partial wave but any arbitrary enhancement introduced
to obtain
 the 4$n$ binding would be in conflict with
the unbound -- or loosely bound~\cite{WG_PRC85_2012}  -- dineutron.
 However,  due to the Pauli principle the effective interaction between dineutrons is mostly repulsive
 and this partial wave does not contribute much in building attraction between the dineutron pairs.

In contrast,   the Pauli principle  does not prevent $P$- and
higher partial waves contributions from increasing the attraction
between a dineutron and another neutron. Moreover $P$-waves have
been a long standing controversy in nuclear
physics~\cite{EpelKam,Wood,Dolesch},  and some few-nucleon
scattering observables (as analyzing powers) would favor stronger
$P$-waves. Nevertheless the discrepancies with scattering data
might be accounted for by  a small  variation of the $nn$ $P$-waves,
of the order of 10\%. In fact, some previous
studies~\cite{Lazauskas_PhD_2003} showed that, in order to bind
the tetraneutron, the  attractive $nn$ $P$-waves should be
multiplied by a factor $\eta\sim4$, rendering the dineutron strongly
resonant in these $P$-waves. One should also notice that if all
the $nn$ $P$-wave interactions are enhanced with the same factor
$\eta$, the dineutron becomes bound  well before the tetraneutron.
In order to create a narrow 4$n$ resonance, a slightly weaker
enhancement is required, but still this enhancement factor remains
considerable,
 $\eta\gtrsim3$. Therefore such a modification strongly
contradicts the nature of the nuclear interaction, which satisfies
rather well isospin conservation.

Finally, as  noted in Ref.~\cite{Pieper2001}, a three-neutron
force might make a key contribution in building the additional
attraction required to generate resonant multineutron clusters. As
we will see in the next section, the presence of an attractive
$T=3/2$  component in the 3$N$ force is clearly suggested in the
studies based on the best $NN$ and $T=1/2$ $3N$ potentials,  which
often underestimate the binding energies of the neutron-rich
systems. Furthermore the contribution of such a force should rise
quickly with the  number of neutrons in the system, and we will
indeed demonstrate this when comparing  3$n$ and 4$n$ systems.

\vskip 0.2cm  
In our  previous
studies~\cite{Lazauskas_PhD_2003,Lazauskas-3n} we have employed
different realistic $NN$ interaction models (Reid93, AV18, AV8$'$,
INOY) in analyzing multineutron systems  and found that they
provide qualitatively the same results. For all these reasons  we
will focus on the modification of the 3$N$ force in the total
isospin $T=3/2$ channel. The main purpose of this work is, thus,
to investigate whether  a resonant tetraneutron state is
compatible with our knowledge of the nuclear interaction, in
particular with the $T=3/2$ 3$N$ force. To this aim we will fix
the $NN$ force with a realistic interaction  and introduce a
simple isospin-dependent $3N$ force acting in both isospin
channels. Its $T=1/2$ part will be adjusted to describe some $A=3$
and $A=4$ nuclear states and the $T=3/2$ part will be tuned until a
$^4n$ resonance is manifested. The exploratory character of this
study, as well as the final conclusions, justify the simplicity of
the phenomenological force adopted here.

\vskip 0.1cm 
The paper is organized as follows. In Section \ref{H}
we will present the $NN$ and $3N$ interactions and the unique
adjustable parameter of the problem. Section \ref{Method} is
devoted to sketching the complex scaling
method~\cite{CSM-ref1,CSM-ref2,CSM-ref3,Ho,Moiseyev} that is used
to solve the $4n$ Schr\"{o}dinger equation under the correct
boundary condition for resonant states. Also we briefly explain
the Gaussian expansion method{~\cite{Kami88,Kame89,Hiya03,
Hiya12FEW,Hiya12PTEP,Hiya12COLD} and the FY equations for solving
the $A=4$ problem. Results obtained are presented and discussed in
the Section \ref{Results}; finally conclusions are drawn in the
last Section \ref{Conclusions}.

\section{Hamiltonian}\label{H}

We start with a general nonrelativistic  nuclear Hamiltonian
\begin{equation}
H=T +\sum_{i<j} V_{ij}^{NN} + \sum_{i<j<k}V_{ijk}^{3N},
\end{equation}
where $T$ is a four-particle kinetic-energy operator,
$V_{ij}^{NN}$ and $V_{ijk}^{3N}$ are respectively two- and
three-nucleon potentials. In this work we use the AV8$'$
version~\cite{AV8P97} of the $NN$ potentials derived by the
Argonne group. This model describes well the main properties of
the $NN$ system and it is relatively easy to handle. The main
properties of this interaction are outlined in the benchmark
calculation of the $^4$He ground state~\cite{Kama01}.

\vskip 0.1cm
As  most  $NN$ forces,  AV8$'$ fails to reproduce binding
energies of the lightest nuclei, in particular those of $^3$H,
$^3$He and $^4$He. A $3N$ interaction is required and we have
therefore supplemented AV8$'$ with a purely phenomenological $3N$
force which is assumed to be isospin-dependent and  given by a sum
of two Gaussian terms:
\begin{equation}\label{V3NT}
V_{ijk}^{3N}=\!\!\sum_{T=1/2}^{3/2}\:
\sum_{n=1}^2 W_n(T)e^{-(r_{ij}^2+r_{jk}^2+r_{ki}^2)/b_n^2} \,
{\cal P}_{ijk}({T})\; .
\end{equation}
where ${\cal P}_{ijk}({T})$ is a projection operator for the total
three-nucleon isospin $T$ state. The parameters of this force --
its strength $W_n$ and range $b_n$ -- are adjusted to reproduce
the phenomenology.

In the case of $T=1/2$  the parameters were fixed in
Ref.~\cite{Hiya04SECOND} when studying the $J^\pi=0^+$ states of 
$^4$He nucleus. They are:
\begin{equation}\label{Para_T1/2}
\begin{array}{rclcl}
W_1(T=1/2)&=& -2.04 \;{\rm MeV},    &&  b_1=4.0\,\; \; {\rm fm}, \cr 
\noalign{\vskip 0.1 true cm}
W_2(T=1/2)&=& +35.0 \;{\rm MeV},   &&  b_2=0.75\, {\rm fm}.
\end{array}
\end{equation}

Using this parameter set, in addition to the AV8$'$ and Coulomb
interaction, one obtains  the following binding energies:
$^3$H=8.41 (8.48) MeV, $^3$He=7.74 (7.72) MeV, $^4$He ($0^+_1$)=
28.44 (28.30) MeV and the excitation energy of
$^4$He$(0^+_2)$=20.25 (20.21) MeV~\cite{Hiya04SECOND}, where the
experimental values are shown in parentheses. Furthermore, this
parameterization allows one to reproduce the observed transition form
factor  $^4{\rm He}(e,e')^4{\rm He}(0^+_2)$ (cf. Fig.~3 of
Ref.~\cite{Hiya04SECOND}). 

Although the $^3$H and $^3$He nuclei
contain in their wave functions a small admixture of isospin $T=3/2$
configurations, these calculations have been performed by
neglecting it, as it is the case in most of the few-nucleon
calculations.

\begin{figure}[b]
\begin{center}
\epsfig{file=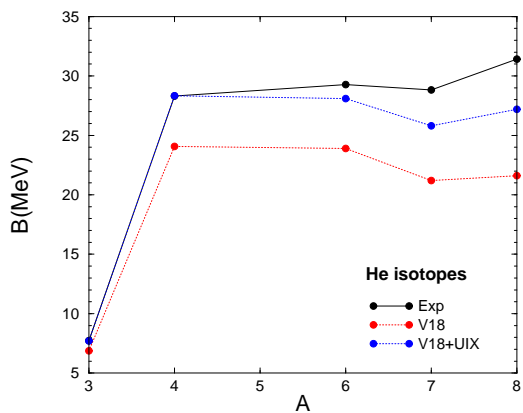,scale=0.4}
\end{center}
\caption{(color online) 
Experimental binding energies of the He isotopes  compared with the 
predictions based on  AV18  and AV18+UIX Hamiltonians. 
The UIX $3N$ force is purely repulsive in the $T=3/2$ channel and 
exhibits attraction only in the $T=1/2$ one.
Displayed values are taken from Table II of Ref.~\cite{Pieper2001}.}
\label{He_Isotopes}
\end{figure}

\vskip 0.1cm
The $4n$ system is only sensitive to the  $T=3/2$ component of the
$3N$ interaction. This component has almost no effect in
proton-neutron symmetric nuclei, but it manifests clearly itself in the
series of He isotopes, where the purely $T=1/2$ $3N$ force,
adjusted to reproduce well the $^4$He, fails to describe the
increasingly neutron-rich He isotopes. This can be illustrated
with the results of the GFMC calculations, Table II of
Ref.~\cite{Pieper2001}, which are displayed in
Fig.~\ref{He_Isotopes}.

\vskip 0.1cm 
This situation was dramatically improved in
Ref.~\cite{Pieper2001}, where several $3\leq A \leq 8$ nuclei were
used to fix the parameters of a new series of spin-isospin
dependent Illinois $3N$ forces (IL1$-$IL5) which reproduce well
the experimental data in Fig.~\ref{He_Isotopes}. It is worth
noting however that, from the results~in~Fig.~\ref{He_Isotopes},
the effect of the $T=3/2$ component of the $3N$ force remains
smaller than the $T=1/2$ component.

\vskip 0.1cm
Throughout the present paper, the attractive strength parameter 
of the $T=3/2$ component,
$W_1(T=3/2)$, will be considered as a free  parameter and
varied in order to  analyze the existence of a possible tetraneutron resonance.
The other parameters retain the same value as 
in the $T=1/2$ case; that is we use:
\begin{equation}\label{Para_T3/2}
\begin{array}{rclcl}
W_1(T=3/2)&=& \;\; {\rm free}, \;      &&  b_1=4.0\,\;  {\rm fm}, \cr
\noalign{\vskip 0.1 true cm}
W_2(T=3/2)&=& +35.0 \;{\rm MeV},   &&  b_2=0.75\, {\rm fm} .
\end{array}
\end{equation}

We will explore in parallel  the effect of such a force on the
$A=4$ nuclei that could be sensitive to the $T=3/2$ component,
that is: $^4$H, $^4$He and $^4$Li, in  states with total isospin
$T=1$ and angular momentum $J^{\pi}=1^-$ and $2^-$.

\section{Computational Method}\label{Method}

Two independent configuration space methods are used in
 solving the four-body problem: the Gaussian expansion
 method~\cite{Kami88,Kame89,Hiya03,Hiya12FEW,Hiya12PTEP,Hiya12COLD}
 is applied to solve Schr\"{o}dinger equation and
 Lagrange-mesh technique applied to solve FY equation. In order to
to simplify boundary conditions related to the four-body problem
in the continuum we employ the complex scaling
method~\cite{CSM-ref1,CSM-ref2,CSM-ref3,Ho,Moiseyev}. These
methods will be briefly sketched in what follows.

\subsection{Complex scaling Method}

In this work, we focus on the possible existence of the narrow
resonant states of $^4n$, which may enhance significantly  $^4n$
production cross section. We employ the complex scaling method
(CSM) in order to calculate resonance positions and widths. The
CSM and its application to nuclear physics problems are
extensively reviewed in Refs.~\cite{Aoyama2006,myo2014} and
references therein. Using the CSM, the resonance energy (its position
and width) is obtained as a stable complex eigenvalue  of the
complex scaled Schr\"{o}dinger equation:
\begin{equation}
[H(\theta) -E(\theta)] \Psi_{JM, TT_z}(\theta)=0 \;
\label{eq:sccsm},
\end{equation}
where $H(\theta)$ is obtained by making the radial transformation
of the four-body Jacobi coordinates (Fig. 1) in $H$ of Eq.~(2.1)
with respect to the common complex scaling angle of $\theta$:
\begin{equation}
r_{\rm c} \to r_{\rm c} \,e^{i \theta}, \;
R_{\rm c} \to R_{\rm c} \,e^{i \theta}, \;
{\rho}_{\rm c} \to    \rho_{\rm c} \,e^{i \theta}
\;\; ({\rm c}={\rm K}, {\rm H})  .
\end{equation}
According to the ABC theorem~\cite{CSM-ref1,CSM-ref2},
the eigenvalues of
Eq.~(\ref{eq:sccsm}) may be separated into three groups:

i) The bound state poles, remain unchanged under the complex
scaling transformation and remain on the negative real axis.

ii) The
cuts, associated with discretized continuum states, are rotated
downward making an angle of $2\theta $ with the real axis.

iii) The resonant poles are independent of parameter $\theta$ and
are isolated from the discretized non-resonant continuum spectrum
lying along the $2\theta$-rotated line when the relation tan\,2$\theta
> -{\rm Im}(E_{\rm res})/{\rm Re}(E_{\rm res})$ is satisfied.
The resonance width is defined by \mbox{$\Gamma=-2\,{\rm Im}(E_{\rm res})$.}

In the next subsection, we shall show, as an example satisfing the above
properties i)-iii),  narrow and
broad $^4n$ resonances and the $4n$ continuum spectrum rotated
into the complex energy plane.

\subsection{Gaussian expansion method}

A great advantage of the CSM is that it allows one to describe the
resonant states using $L^2$-integrable wave functions. Therefore,
the Gaussian expansion method
(GEM)~\cite{Kami88,Kame89,Hiya03,Hiya12FEW,Hiya12PTEP,Hiya12COLD}
has been successfully applied in conjunction with the CSM in 
nuclear few-body calculations~\cite{Aoyama2006,myo2014,PPNP} as
well as in recent three- and four-body calculations by two
authors (E.H. and M.K.) of the present
manuscript~\cite{Ohtsubo2013,Hiya13H6L,Hiya14He7L}.

In order to expand the system's wave function $\Psi_{JM,
TT_z}(\theta)$ we employ the Gaussian basis functions of the same
type as those used in the aforementioned references. An isospin
rather than a neutron-proton (particle) basis is used to
distinguish between different nuclear charge states $^4n$, $^4$H,
$^4$He and $^4$Li. In the GEM approach, the four-nucleon wave
function is written as a sum of the component functions in the K-
and H-type Jacobi coordinates (Fig.~\ref{fig:he4jacobi}),
employing the $LS$ coupling scheme:
\begin{eqnarray}
\!\!\! \Psi_{JM, TT_z}(\theta) =
\!\sum_{\alpha}  C^{({\rm K})}_{\alpha}(\theta)
\Phi^{({\rm K})}_\alpha +
\!\sum_{\alpha}  C^{({\bf H})}_{\alpha}(\theta)
\Phi^{({\rm H})}_\alpha  ,
\label{Psi-expansion}
\end{eqnarray}
where the antisymmetrized four-body basis functions $\Phi^{({\rm
K})}_\alpha$ and $\Phi^{({\rm H})}_\alpha$ (whose suffix $JM,TT_z$
are dropped for simplicity) are described by
\begin{eqnarray}
 \Phi^{({\rm K})}_\alpha\!\! &=& \!\!       {\cal A}
 \left\{
\Big[ \big[ [\phi_{nl}^{({\rm K})}({\bf r}_{\rm K})
          \varphi_{\nu\lambda}^{({\rm K})}
          (\rhovec_{\rm K})]_\Lambda \;
       \psi_{NL}^{({\rm K})}({\bf R}_{\rm K}) \big]_{I}
     \right.    \nonumber \\
&& \quad \times      \big[ [\chi_s(12)
        \chi_{1/2}(3)]_{s'}
        \chi_{1/2}(4)
          \big]_{S}    \Big]_{JM}   \nonumber \\
&&\quad \times   \left.      \big[ [\eta_t(12)
        \eta_{1/2}(3)]_{t'}
        \eta_{1/2}(4)
          \big]_{TT_z} \right\} ,
\label{eq:Psi-K}  \\
\Phi^{({\rm H})}_\alpha \!\!& =& \!\!     {\cal A}
 \left\{
\Big[  \big[[\phi_{nl}^{({\rm H})}({\bf r}_{\rm H})
         \varphi_{\nu\lambda}^{({\rm H})}
           (\rhovec_{\rm H})]_\Lambda \;
        \psi_{NL}^{({\rm H})}({\bf R}_{\rm H}) \big]_{I}
     \right.     \nonumber \\
&&\quad \times     \big[ \chi_s(12)\chi_{s'}(34)
          \big]_{S}    \Big]_{JM}    \nonumber \\
&&\quad \times   \left.
          \big[ \eta_t(12)\eta_{t'}(34)
          \big]_{TT_z}  \right\}  ,
          \label{eq:amp}
\end{eqnarray}
with $\alpha \equiv \{nl,\nu \lambda,\Lambda,NL,I,s,s',S,t,t' \}$.
$\cal{A}$ is the four-nucleon antisymmetrizer.
The parity of the wave function is given by $\pi=(-)^{l+\lambda+L}$.
The $\chi$'s and $\eta$'s are the spin and isospin functions,
respectively.  The spatial basis functions $\phi_{nl}({\bf r})$,
 $\varphi_{\nu\lambda} (\rhovec)$ and $\psi_{NL}({\bf R})$
are taken to be Gaussians multiplied by spherical harmonics:
\begin{eqnarray}
&&\phi_{nlm}({\bf r}) =
N_{nl}\,r^l\:e^{-(r/r_n)^2}\:
Y_{lm}({\widehat {\bf r}})   \;,\nonumber \\
&&\varphi_{\nu \lambda \mu}(\rhovec)=
N_{\nu \lambda}\,\rho^\lambda\:e^{-(\rho/\rho_\nu)^2}\:
Y_{\lambda \mu}({\widehat {\rhovec}})
\;, \\ 
\label{eq:4gauss}
&&\psi_{NLM}({\bf R}) =
N_{NL}\,R^L\:e^{-(R/R_N)^2}\:Y_{LM}({\widehat {\bf R}})
\;.  \nonumber
\end{eqnarray}

It is important to postulate that the Gaussian ranges 
lie in geometric progression (for the reason, see Sec.~IIB of the first
paper of Ref.~\cite{Hiya12COLD}):
\begin{eqnarray}
&& r_n=r_1\, a^{n-1}\;\quad
\quad \:(n=1-n_{\rm max})\;, \nonumber \\
&&\rho_\nu=
\rho_1\, \alpha^{\nu-1}\;\quad \quad
(\nu=1-\nu_{\rm max})\;,\\ 
&&R_N=R_1\, A^{N-1}\;\quad
(N=1-N_{\rm max})\;. \nonumber
\end{eqnarray}
Such a choice of basis functions is suitable for simultaneous
description of
 short-range correlations and long-range asymptotic behavior
 (for example, see Refs.~\cite{Kami88,Kame89,Hiya03,
Hiya12FEW,Hiya12PTEP,Hiya12COLD,Ohtsubo2013}).

\begin{figure}[t]
\begin{center}
\epsfig{file=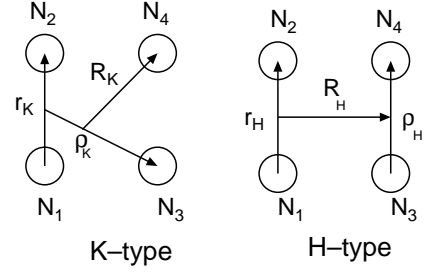,scale=0.5}
\end{center}
\caption{Four-nucleon  Jacobi coordinates of K-type and H-type
configurations. .} \label{fig:he4jacobi} 
\end{figure}

\vskip 0.1cm
The eigenenergy $E(\theta)$ and the expansion coefficients
in Eq.~(\ref{Psi-expansion}) are obtained by
diagonalizing the Hamiltonian $H(\theta)$
with the basis functions (\ref{eq:Psi-K}) and (\ref{eq:amp}).
In the following calculations, satisfactory convergence was obtained
within $l, L, \lambda \leq 2$ (cf. an example of rapid convergence
in the binding energies of $^3$H and $^3$He
with realistic $NN$ and $3N$ interactions
is presented in Refs.~\cite{Kame89,Hiya03}).
We note that, in the GEM framework,
contrary to the truncation in the wave function,
the interaction is included without partial-wave decomposition
(no truncation in the angular-momentum space);
this makes the convergence rapid
(cf. discussion on this point in \S 2.2 of Ref.~\cite{Gibson93}).

\begin{figure}[t]
\begin{center}
\begin{minipage}[t]{9.0 cm}
\epsfig{file=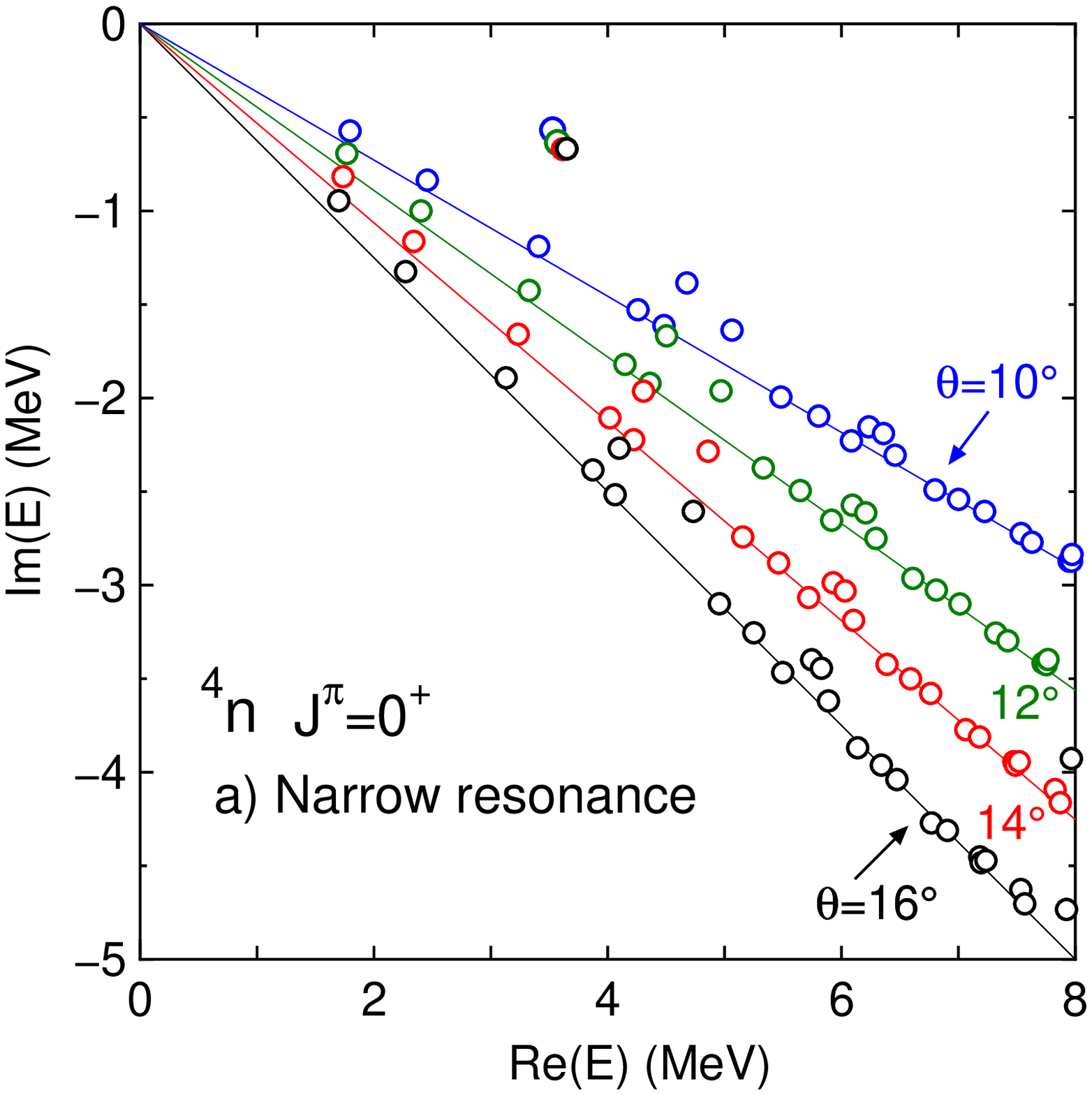,scale=0.4}
\end{minipage}
\vskip 0.2cm
\hspace{\fill}
\begin{minipage}[t]{9.0 cm}
\epsfig{file=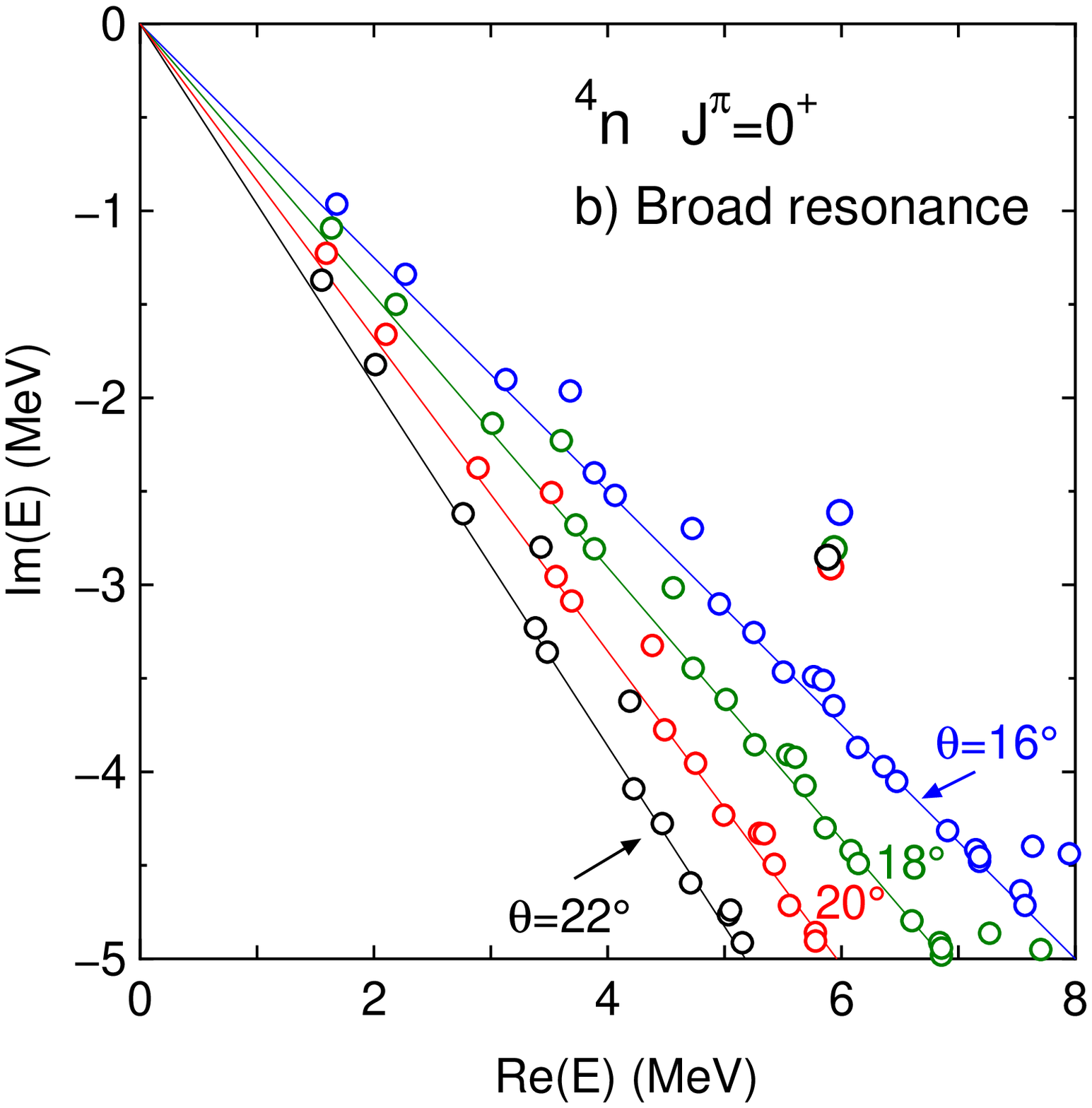,scale=0.4}
\end{minipage}
\end{center}
\caption{ (color online)   Dependence of the eigenenergy distribution
on the complex scaling angle $\theta$ for the  $^4n$ system with
$J^\pi=0^+$. Two different cases are considered a) presence of a
narrow resonance at $E_{\rm res}=3.65-0.66i$ MeV for
$W_1(T=3/2)=-28$ MeV and b) presence of a broad resonance at
$E_{\rm res}=5.88-2.85i$ MeV for $W_1(T=3/2)=-21$ MeV. }
\label{fig:pole-example}
\end{figure}

\vskip 0.1cm  In Fig.~\ref{fig:pole-example} we present two
typical applications of the GEM in diagonalizing the complex-scaled
Hamiltonian $H(\theta)$ that describes the $^4n$ system. The two
panels of  this figure  display  the \mbox{$\theta$-dependence} of
the eigenenergies for the cases where the $^4n$ system possesses a) a
narrow resonance and b) a broad resonance with 
$\Gamma \! = -2\,{\rm Im}(E_{\rm res}) \! \sim \! 6$ MeV. 
The resonance pole position
converges when increasing the complex scaling angle $\theta$,  and
the pole becomes well isolated from the four-body
continuum along the $2\theta$-line. To obtain the result in
Fig.~\ref{fig:pole-example}, some 14000 antisymmetrized four-body
basis functions were needed.

\subsection{Faddeev-Yakubovsky equations}

The FY equations use a very similar representation of the system's wave
function as the one employed by the GEM and presented in a previous
section.
The FY equations are formulated in terms of wave function components,
  which are very similar to the ones expressed in Eqs.~(\ref{eq:Psi-K})
and~(\ref{eq:amp}), namely:
\begin{eqnarray}
 {\cal F}^{({\rm K})}_\alpha\!\! &=& \!\!
 \!\sum_{\alpha_K}C_{\alpha_K}
\Big[ \big[ [\phi_{nl}^{({\rm K})}({\bf r}_{\rm K})
          \varphi_{\nu\lambda}^{({\rm K})}
          (\rhovec_{\rm K})]_\Lambda \;
       \psi_{NL}^{({\rm K})}({\bf R}_{\rm K}) \big]_{I}
   \nonumber \\
&\times&     \!\!     \big[ [\chi_s(12)
        \chi_{1/2}(3)]_{s'}
        \chi_{1/2}(4)
          \big]_{S}    \Big]_{JM}   \nonumber \\
&\times&  \!\!
     \big[ [\eta_t(12)
        \eta_{1/2}(3)]_{t'}
        \eta_{1/2}(4)
          \big]_{TT_z}   ,   \\
{\cal F}^{({\rm H})}_\alpha \!\!& =&
\!\!\sum_{\alpha_H}C_{\alpha_H}
\Big[  \big[[\phi_{nl}^{({\rm H})}({\bf r}_{\rm H})
         \varphi_{\nu\lambda}^{({\rm H})}
           (\rhovec_{\rm H})]_\Lambda \;
        \psi_{NL}^{({\rm H})}({\bf R}_{\rm H}) \big]_{I}
         \nonumber \\
&\times&    \!\!    \big[ \chi_s(12)\chi_{s'}(34)
          \big]_{S}    \Big]_{JM}    
          \big[ \eta_t(12)\eta_{t'}(34)
          \big]_{TT_z}    .
\end{eqnarray}

The major difference in the expansions of Eqs.~(\ref{eq:Psi-K})
and~(\ref{eq:amp}) employed for the GEM is that the FY components
are not  straightforwardly antisymmetrized. Symmetry of the total
wave function is enforced by the FY equations whose components are
subject to, namely:
\begin{small}
\begin{eqnarray*}
\nonumber \left( E-H_{0}-V_{12}\right) {\cal F}^{({\rm K})}_\alpha\!&-&\!V_{12}(P^{+}+P^{-})\left[ (1+Q){\cal F}^{({\rm K})}_\alpha+{\cal F}^{({\rm H})}_\alpha\right]\\
&&=\frac{1}{3}V_{123}\, \Psi_{JM, TT_z}, \label{FY_drive1}  \nonumber \\
\left( E-H_{0}-V_{12}\right) {\cal F}^{({\rm H})}_\alpha \!&-&\! V_{12}\tilde{P}\left[ (1+Q){\cal F}^{({\rm K})}_\alpha+{\cal F}^{({\rm H})}_\alpha\right] =0,      \label{FY_drive2}
\end{eqnarray*}
\end{small}
$\!\!$where $P^-=P_{23}P_{12}$, $P^+=P_{12}P_{23}$, $Q=P_{34}$ and
$\tilde{P}=P_{13}P_{24}$ are particle permutation operators.

\vskip 0.1cm 
In terms of the FY components the total system wave function
is obtained as
\begin{small}
\begin{eqnarray}
\Psi_{JM, TT_z}(\theta)\!&=&\!\!\sum_{\alpha}[1+(1+P^{+}+P^{-})Q]
(1+P^{+}+P^{-}){\cal F}^{({\rm K})}_\alpha \nonumber  \\
&+&\!\sum_{\alpha}(1+P^{+}+P^{-})\tilde{P}{\cal F}^{({\rm H})}_\alpha .
\end{eqnarray}
\end{small}

It is straightforward to apply the complex scaling operation to the FY
equations, as has been demonstrated for the Schr\"{o}dinger Eq.
(\ref{eq:sccsm}). For a more detailed explanation see one of our
previous papers~\cite{Lazauskas-4n,PPNP}.

The transformed FY equations are solved using standard techniques,
developed in~\cite{Lazauskas_PhD_2003,Lazauskas-4n} and references
therein. The radial dependence of the complex-scaled FY components
${\cal F}^{({\rm K})}_\alpha$ and ${\cal F}^{({\rm H})}_\alpha$ is
expanded in a Lagrange-Laguerre basis and the system of
integro-differential equations is transformed into a linear
algebra problem by using the Lagrange-mesh method~\cite{Baye}.
Lagrange-mesh of $\sim(20-25)^3$ points is required to describe
accurately the radial dependence of the FY components. The FY
equations converge considerably slower in partial-waves compared
to the Schr\"{o}dinger equation case using the GEM. When solving 
the FY equations we
include partial-waves with angular momenta $max(l, L,
\lambda)\leq7$. Slow convergence of these calculations is related
to the $3N$ force terms we employed in this work, whose
contribution turns out to be unnaturally large. The FY equations are
not very well formulated to handle these terms.

\section{Results and Discussion}\label{Results}

The recent experiment, providing evidence of 
the possible existence of a resonant tetraneutron,
reported some structure at $E\!=\!0.83
\!\pm\! 0.65 ({\rm stat.})\! \pm\! 1.25 ({\rm sys.})$ MeV, measured with
respect to the $4n$ breakup threshold with an estimated upper
limit width {$\Gamma\!=\!2.6$ MeV\mbox{~\cite{Kisamori_PhD,Kisamori_PRL}}. 
This experiment, as
well as others  reporting a positive tetraneutron signal 
\cite{Marques:2001wh,Marques:2005vz,Chulkov_NPA750_2005} were
not able to extract information on the spin-parity of the observed events.
Our first task  is therefore to determine the most favorable angular 
momentum states to accommodate a tetraneutron.

\subsection{4n bound state}\label{4n_bs}

For this purpose, we calculate a critical strength of the attractive 3$N$
force $W_1(T=3/2)$, defined by Eq.~(\ref{V3NT}), to make different
$4n$ states bound at $E=-1.07$ MeV. This energy corresponds to the
lowest  value compatible with the RIKEN
data~\cite{Kisamori_PRL}. The calculated results, denoted as
$W_1^{(0)}(T\!=\!3/2)$,  are given in Table~\ref{table:critical-strength}.

\begin{table} [htb] \begin{center}
\caption{Critical strength   $W_1^{(0)}(T=3/2)$ (MeV)  of the phenomenological 
$T=3/2$ $3N$ force
required to bind the $4n$ system at $E=-1.07$ MeV, the
lower bound of the experimental value~\cite{Kisamori_PRL}, 
for different states as well
as the probability (\%) of their four-body  partial waves.}
\label{table:critical-strength}
\begin{tabular}{ccccccc}
\hline
\hline
\noalign{\vskip 0.1 true cm}
 $J^{\pi}$  & $0^+$  &$1^+$  &$2^+$  &$0^-$  &$1^-$  &$2^-$ \\ \noalign{\vskip 0.1 true cm} \hline \noalign{\vskip 0.15 true cm} $W_1^{(0)}(T\!=\!\frac{3}{2})$ & $-36.14$ & $-45.33$ & $-38.05$ &
  $-64.37$ & $-61.74$ & $-58.37$  \\
\noalign{\vskip 0.15 true cm}
\hline
\noalign{\vskip 0.15 true cm}
$S$-wave & $93.8$ & $0.42$  & $0.04$ &
           $0.07$ & $0.08$ & $0.08$  \\ $P$-wave & $5.84$ & $98.4$  & $17.7$ &
          $99.6$ & $97.8$ & $89.9$  \\
$D$-wave & $ 0.30$ & $1.08$  & $82.1$ &
          $0.33$ & $2.07$ & $9.23$  \\
$F$-wave & $ 0.0$ & $0.05$  & $0.07$ &
           $0.0$ & $0.10$ & $0.74$  \\
\noalign{\vskip 0.15 true cm}
\hline
\end{tabular}
\end{center}
\end{table}

\vskip 0.1cm As one can see from this table, the smallest critical
strength is $W_1^{(0)}(T\!=\!3/2)=-36.14$ MeV and corresponds to
the $J=0^+$ state.  It is consistent with a result reported in
Ref.~\cite{Lazauskas-4n}, where the tetraneutron binding was forced
using an artificial four-body force in conjunction with the Reid93 $nn$
potential. The next most favorable configuration is established to be
a $2^+$ state, which is bound by $1.07$ MeV for a 3NF strength of
$W_1^{(0)}(T\!=\!3/2)$. The calculated level ordering is
$J^{\pi}=0^+, 2^+, 1^+,  2^-, 1^-, 0^-$. The level ordering
calculated in Ref.~\cite{Lazauskas-4n} is $J^{\pi} =0^+, 1^+, 1^-,
2^-, 0^-, 2^+$. These differences are related to the different
binding mechanism of the four-nucleon force used in
Ref.~\cite{Lazauskas-4n}.

\vskip 0.1cm It should be noted that,  in comparison with
$W_1(T\!=\!1/2)=-2.04 $ MeV established for the $T=1/2$ $3N$
force, we need extremely strong $T=3/2$ attractive term to make
the 4$n$ system weakly bound; when the $J=0^+$ state is at
$E=-1.07$ MeV with $W_1(T\!=\!3/2)=-36.14$ MeV, the expectation
values of the kinetic energy, $NN$  and $3N$ forces are  $+67.0,
-38.6$ and $-29.5$ MeV, respectively. We see that the expectation
value of the $3N$ potential  is almost as large as 
that of $NN$ potential. The
validity of this strongly attractive $T=3/2$ $3N$ force will be
discussed after presenting results for $4n$ resonant states.

\subsection{4n resonances}\label{4n_res}

\vskip 0.2cm After determining critical strength of $W_1(T=3/2)$
required to bind the tetraneutron we gradually release this parameter
letting the 4$n$ system to move into the  continuum. In this way we
follow complex-energy trajectory of the $^4n$ resonances for
$J=0^+, 2^+$ and $2^-$ states. We remind the readers that these trajectories
are controlled by a single parameter $W_1(T=3/2)$, whereas other
parameters remain  fixed at the values given in Eq.(\ref{Para_T1/2}) 
and Eq.(\ref{Para_T3/2}).

\begin{figure}[b]
\begin{center}
\begin{minipage}[b]{9.0 cm}
\epsfig{file=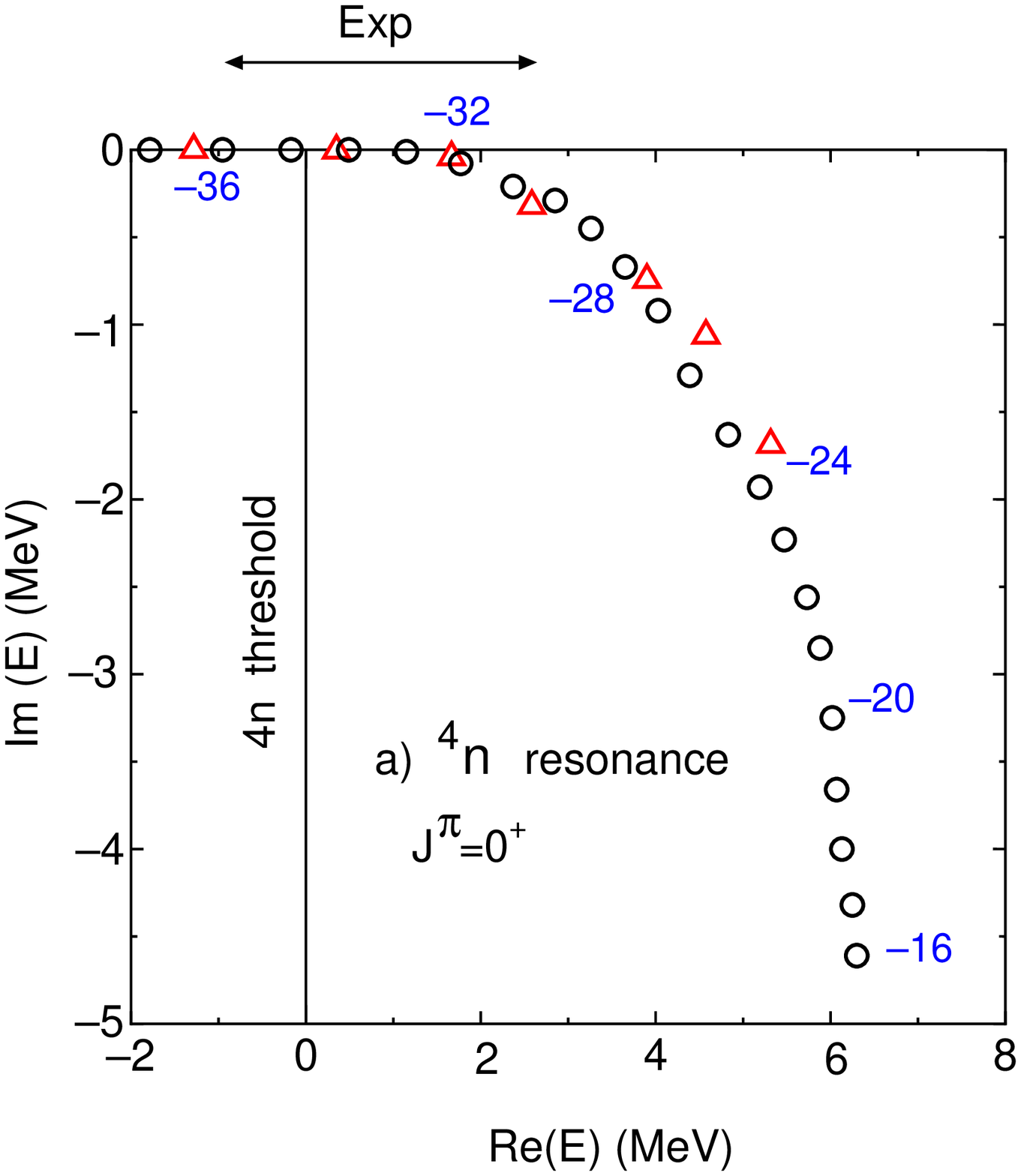,scale=0.4}
\end{minipage}
\vskip 0.2cm
\hspace{\fill}
\begin{minipage}[b]{9.0 cm}
\epsfig{file=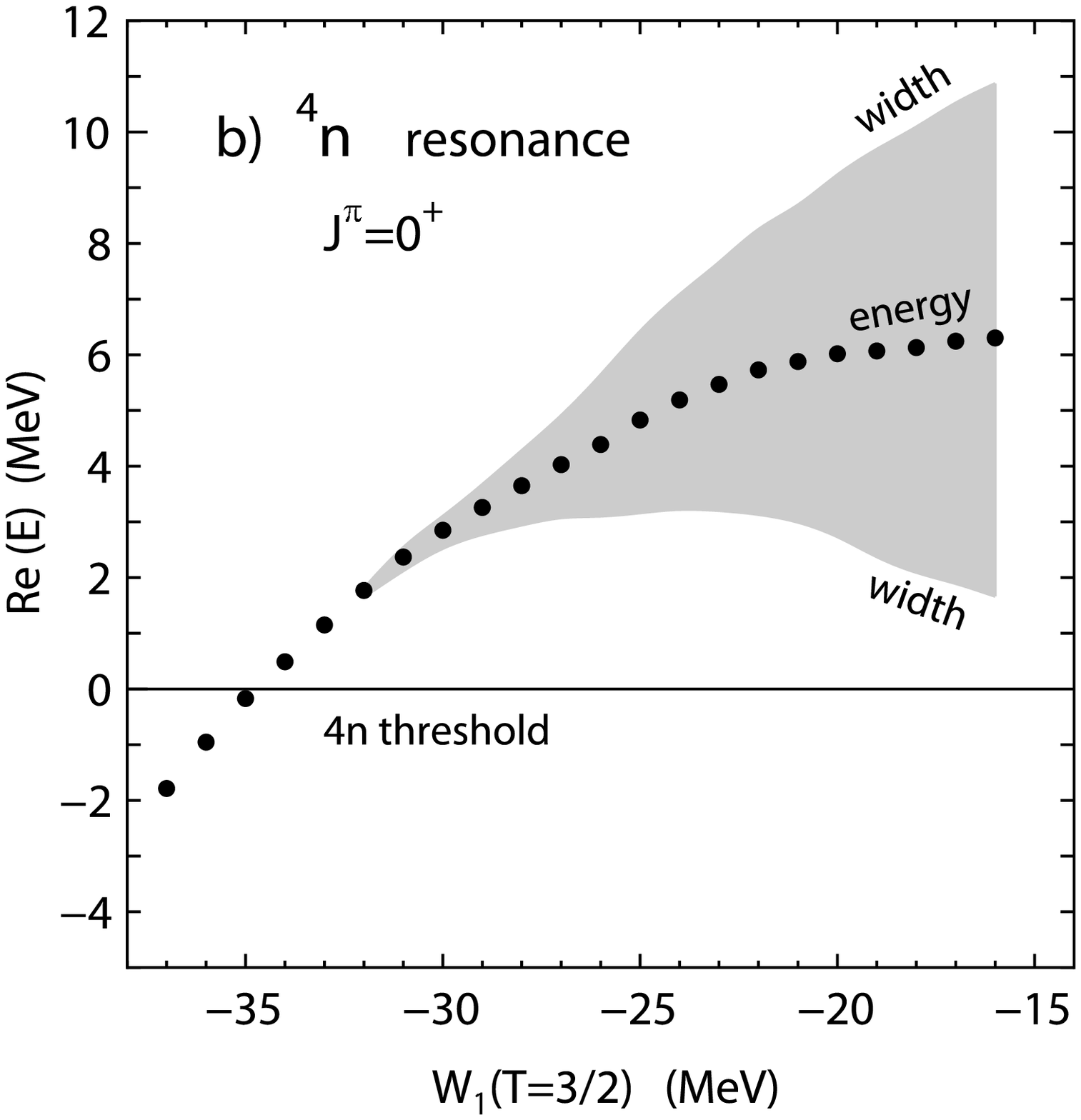,scale=0.40}
\end{minipage}
\end{center}
\caption{a) Tetraneutron resonance trajectory for the $J^\pi=0^+$ state. 
The circles correspond to resonance positions for the AV8$'$ and the
triangles INOY04'(is-m) potential~\cite{Dolesch}. Parameter
$W_1(T=3/2)$ of the additional 3NF was changed from $-37$ to $-16$ MeV 
in steps of 1 MeV
for calculations based on AV8$'$  and from $-36$ to $-24$ MeV 
in steps of 2 MeV for INOY04'(is-m).
To guide the eye the resonance region suggested by the 
measurement~\cite{Kisamori_PRL} is
indicated by the arrow at the top.
b) The same contents as in the upper panel figure (AV8$'$), but where the 
resonance energy (closed circles) and width (shadowed area) are 
represented as a function of the $W_1(T=3/2)$ parameter.
}
\label{fig:nnnn-trajectry}
\end{figure}

In Fig.~\ref{fig:nnnn-trajectry}a, we display the $^4n$ S-matrix
pole (resonance) trajectory for $J=0^+$ state by reducing the
strength parameter
from $W_1(T\!=\!3/2)=-37$ to $-16$ MeV in step of $1$ MeV.
We were unable to continue the resonance trajectory beyond the $W_1
(T=3/2) =-16$ MeV value with the CSM, the resonance becoming too broad to
be separated from the non-resonant continuum.
To guide the eye, at the top of the same figure, we presented an arrow
to indicate the $^4n$ real energy range suggested by the recent
measurement~\cite{Kisamori_PRL}.
In that range the maximum value of the calculated
decay width $\Gamma$ is 0.6 MeV, which is to be compared
with the observed upper limit width $\Gamma=2.6$ MeV.
 In Fig.~\ref{fig:nnnn-trajectry}b the
contents of Fig.~\ref{fig:nnnn-trajectry}a are illustrated in a
different manner to display explicitly the resonance energy and
width versus $W_1(T\!=\!3/2)$. The real energy of the resonances
reaches its maximum value of ${\rm Re}(E_{\rm res})\sim\! 6$ MeV.
Once its real energy maximum is reached the width starts quickly
increasing as the strength $W_1(T\!=\!3/2)$ is further reduced.

\begin{figure}[t]
\vskip 0.5cm
\begin{center}
\epsfig{file=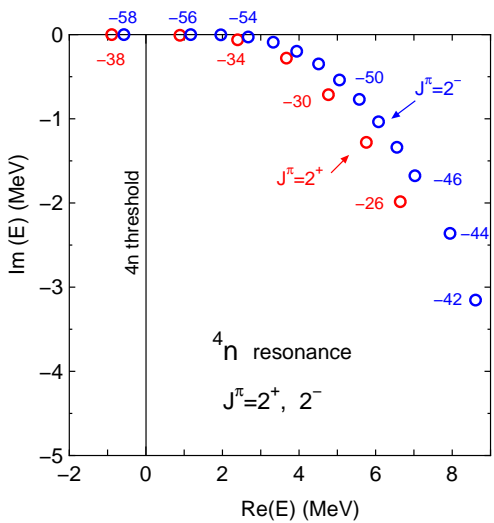,scale=0.4}
\end{center}
\caption{(Color online)
Tetraneutron  resonance trajectories for $J^\pi=2^+$ and $2^-$ states for $W_1(T=3/2)$ values from $-38$ to $-26$ MeV and
from $-58$ to $-42$ MeV, respectively.
}
\label{fig:nnnn-width-avoid}
\end{figure}

\vskip 0.1cm
As was expected, based on our experience from previous studies on
multineutron systems~\cite{Lazauskas_PhD_2003,Lazauskas-3n},
tetraneutron trajectory turns out to be independent of 
the  $NN$ interaction model, provided this model reproduces well the 
$NN$ scattering data. 
To illustrate this feature we have  calculated the 
$^4n$ resonance trajectory for
$J=0^+$state using the INOY04(is-m) $NN$ model~\cite{Dolesch}. This
realistic interaction  strongly differs from the other
ones in that it contains a fully phenomenological and a strongly
non-local short range part in addition to the typical local long
range part based on one pion-exchange. 
Furthermore, this model reproduce the triton and alpha-particle 
binding energies without any 3NF contribution. 
Finally, $P$-waves of this interaction are slightly modified 
in order to match better
the low energy scattering observables in the $3N$ system. Regardless of 
the mentioned qualitative differences of the INOY04(is-m) interaction with
respect to the AV8$'$ one, the results for the $^4n$ resonance
trajectory are qualitatively the same and demonstrate only minor
quantitative differences. These results are displayed 
in Fig.~\ref{fig:nnnn-trajectry}a.

\vskip 0.1cm In Fig.~\ref{fig:nnnn-width-avoid}, we present calculated $^4n$
resonance trajectories for $2^+$ and $2^-$ states. The $J=2^+$ state is
the next most favorable configuration to accommodate a bound
tetraneutron, whereas $J=2^-$ state is the most favorable negative
parity state; see Table~\ref{table:critical-strength}. The
trajectory of the $2^+$ state is very similar to that of the $0^+$ state.
 On the other hand in order to bind  or even to produce
a resonant $J=2^-$  state, in the
region relevant for a physical observation, the attractive
three-nucleon force term $W_1(T\!=\!3/2)$ should be almost twice
as large as the one for $J=0^+$  state. The strength of
$W_1(T\!=\!3/2)$ required to produce a resonant $4n$ system in any
configuration  producing a pronounced experimental signal,
is one order of magnitude larger than the value of 
$W_1(T\!=\!1/2) (-2.04$ MeV) required to
reproduce the binding energies of $^3$H, $^3$He and $^4$He nuclei.

\vskip 0.1cm In order to prove or disprove the possible existence  of 
the tetraneutron resonances,
one sould  consider the validity of the strongly attractive  
3$N$ force in the isospin $T=3/2$ channel. 

As pointed out in Sec.~II, the GFMC calculation for $3\leq A \leq 8$ 
suggested the existence a 3NF with a $T=3/2$  component  weaker  than
the $T=1/2$ one. From the same study it follows that the binding
energies of neutron-rich nuclei are described without notable 
contribution of the $T=3/2$ channel in 3NF. A similar conclusion was reached  
in neutron  matter calculations, where the expectation values of the $T=3/2$ 
force are always smaller than the $T=1/2$ ones \cite{VNNN}.
One should mention that the parametrization of the
phenomenological 3NF adapted in this study is very appropriate 
for dilute states, like the expected tetraneutron resonances.
The attractive 3NF term has a larger range than the one allowed by 
pion-exchange. Moreover tetraneutron states, unlike compound
$^4$He or $^3$H ground states, do not feel contributions of the 
short-ranged repulsive term of the 3$N$ force.}

Thus, we find  no physical justification for the fact that the
$T\!=\!3/2$ term should be one order of magnitude more attractive 
than the $T=1/2$ one, as is required to generate
tetraneutron states compatible with the  ones claimed in the recent  
experimental data~\cite{Kisamori_PRL}.

\subsection{T=1 states in $^4$H, $^4$He and $^4$Li}\label{A=4}

\begin{table} [b]
\begin{center}
\caption{Observed energies $E_R$ and widths $\Gamma$ (in MeV)
of the $J^\pi=2_1^-$ and $1_1^-$ states
in $^4$H,  $^4$He $(T=1)$ and $^4$Li, \mbox{$E_R$ being  measured} from the
$^3$H$+n$, $^3$H$+p$ and $^3$He$+p$ thresholds,
respectively~\cite{Tilley1992}.
}
\label{table:Exp-A=4}
\begin{tabular}{ccccccc}
\noalign{\vskip 0.1 true cm}
\hline
\hline
\noalign{\vskip 0.1 true cm}
  & & $^4$H & & $^4$He $(T=1)$   & & $^4$Li \\
\noalign{\vskip 0.1 true cm}
 $J^{\pi}$  & & $E_R \, (\Gamma)$  & & $E_R \, (\Gamma)$
& & $E_R \, (\Gamma)$  \\
\noalign{\vskip 0.1 true cm}
\hline \noalign{\vskip 0.15 true cm}
$2_1^-\;$ & $\;$ & 3.19 (5.42)  & $\;$ & 3.52 (5.01) &$\;$  & 4.07 (6.03) \\
$1_1^-\;$ & & 3.50 (6.73)  & & 3.83 (6.20) & & 4.39 (7.35) \\
\noalign{\vskip 0.15 true cm}
\hline
\end{tabular}
\end{center}
\end{table}

\begin{figure}[b]
\begin{center}
\begin{minipage}[b]{9.0 cm}
\epsfig{file=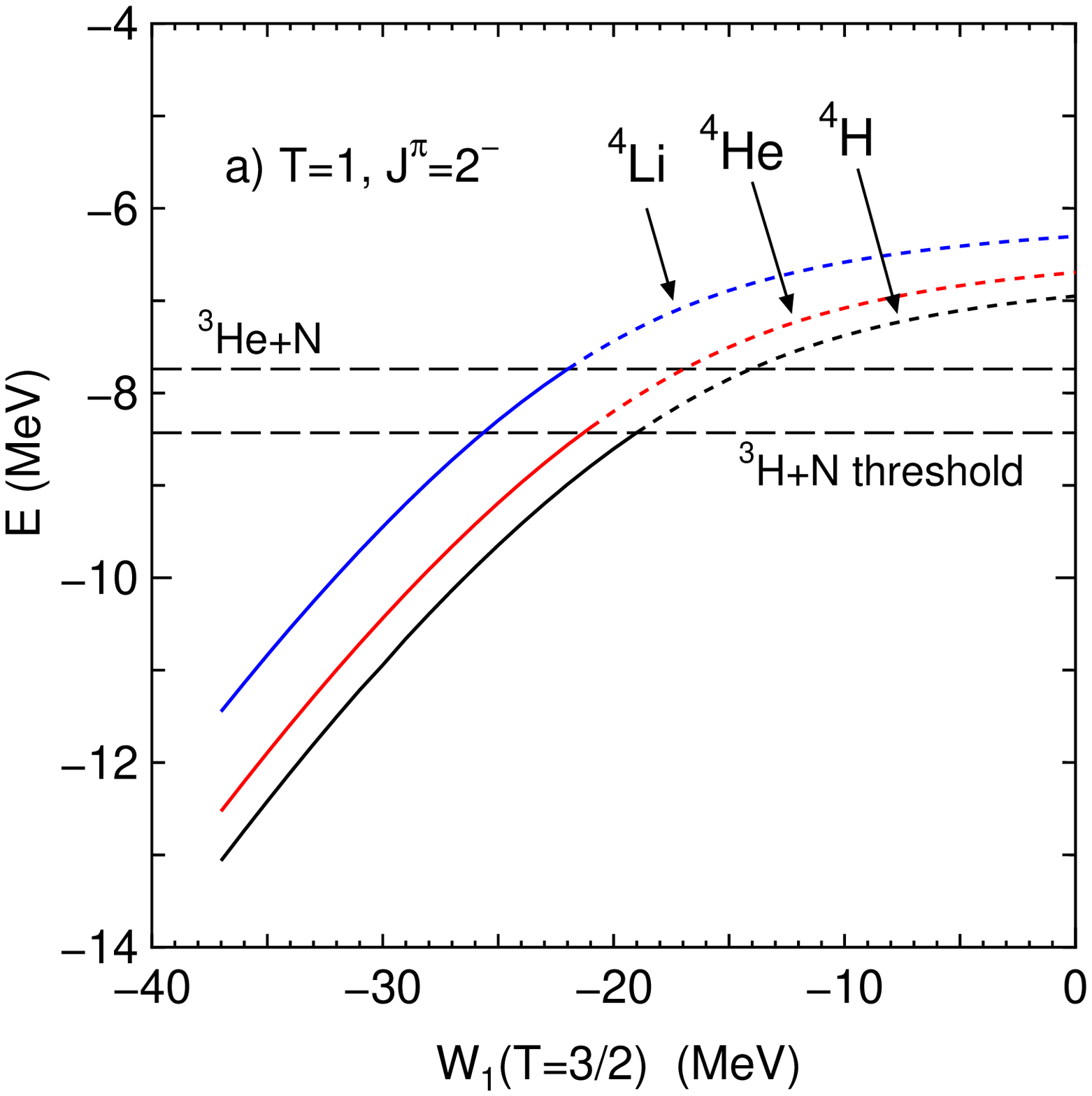,scale=0.4}
\end{minipage}
\vskip 0.2cm
\hspace{\fill}
\begin{minipage}[b]{9.0 cm}
\epsfig{file=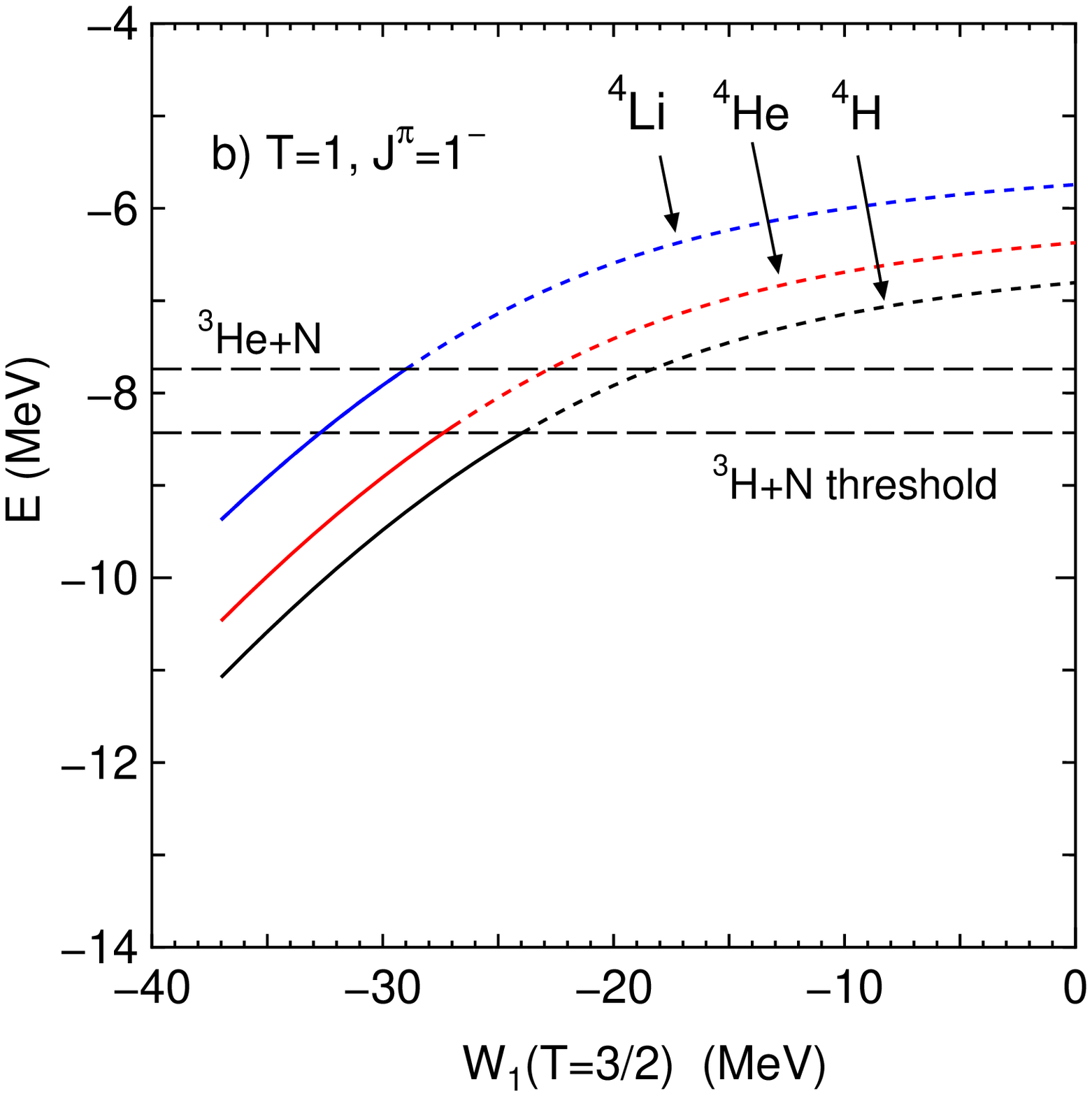,scale=0.4}
\end{minipage}
\end{center}
\caption{(color online) a) Calculated energies of the lowest
$T=1, J^\pi=2^-$ states in $^4$H, $^4$He and $^4$Li
with respect to the strength of
$T=3/2$  $3N$ force, $W_1(T=3/2)$.
\mbox{b) The same} but for $T=1, J^\pi=1^-$ states.
The horizontal dashed lines show the $^3{\rm He}+N$
and $^3{\rm H}+N$ thresholds.
The solid curve below the corresponding threshold
indicates the bound state, while
the dotted curve above the threshold stands approximately for the
resonance obtained by the diagonalization of $H(\theta=0)$
with the $L^2$ basis functions.
}
\label{fig:W1-dependence}
\end{figure}

In the following we would like to investigate the consequences of
a strongly attractive 3NF component in the isospin $T\!=\!3/2$
channel. It is clear that such a force will have the most dramatic
effect on nuclei with a large isospin number, i.e. neutron (or
proton) rich ones as well as on infinite neutron matter.
Nevertheless this includes mostly nuclei with $A>4$, not within
our current scope.
Still we will investigate the effect on other well known states of
$A=4$ nuclei, namely negative parity, isospin $T=1$ states of
$^4$H, $^4$He  and $^4$Li. These structures represent broad
resonances~\cite{Tilley1992} (see Table~\ref{table:Exp-A=4})
established in nuclear collision experiments. Calculated energies
of those states are shown in Fig.~\ref{fig:W1-dependence} with
respect to increasing $W_1(T=3/2)$ from $-37$ to \mbox{0 MeV}. The
solid curve below the corresponding threshold indicates a bound
state, whereas the dotted curve above the threshold stands
approximately for the resonant state  obtained within a bound
state approximation, that is, by diagonalizing $H(\theta=0)$ with
the $L^2$ basis functions (\ref{eq:Psi-K}) and (\ref{eq:amp}).

\vskip 0.1cm As demonstrated in Fig.~\ref{fig:W1-dependence},
values of an attractive 3NF term in the range of $W_1(T\!=\!3/2)
\simeq -36$ to $-30$ MeV, which is compatible with a reported
$^4n$ resonance region in Ref.~\cite{Kisamori_PRL}, gives rise to
the appearance of bound $J=2^-$ and $J=1^-$ states in $^4$H,
$^4$He($T=1$) and $^4$Li nuclei. Unlike observed in the collision
experiments, these states become stable with respect to the $^3$H
$(^3$He$)+N$ decay channels. This means that the present
phenomenological $W_1(T=3/2)$ is too attractive to reproduce
low-lying states of  $^4$H, $^4$He ($T=1$) and $^4$Li.

\vskip 0.1cm In contrast, it is interesting to see the energy of $4n$
system when we have just unbound states for  $^4$H, $^4$He ($T=1$)
and $^4$Li in Fig.~\ref{fig:W1-dependence}a. Use of
$W_1(T=3/2)=-19$ MeV gives rise to an unbound state with $J=2^-$ in
$^4$H with respect to disintegration into  $^3{\rm H}+N$.
However, using this strength of $W_1(T=3/2)$, we have already a
very broad $^4n$ resonant state at Re($E_{\rm res})=6$ MeV with
$\Gamma=7.5$ MeV, see Fig.~\ref{fig:nnnn-trajectry}a, which is
inconsistent with the recent experimental
claim~\cite{Kisamori_PRL} of a resonant $^4n$. Moreover the value of
$W_1(T=3/2)$ that reproduces the observed broad resonance data for the
$2^-$ state in $^4$H should be much less attractive than $-19$ MeV.

\begin{figure}[b]
\vskip 0.5cm
\begin{center}
\epsfig{file=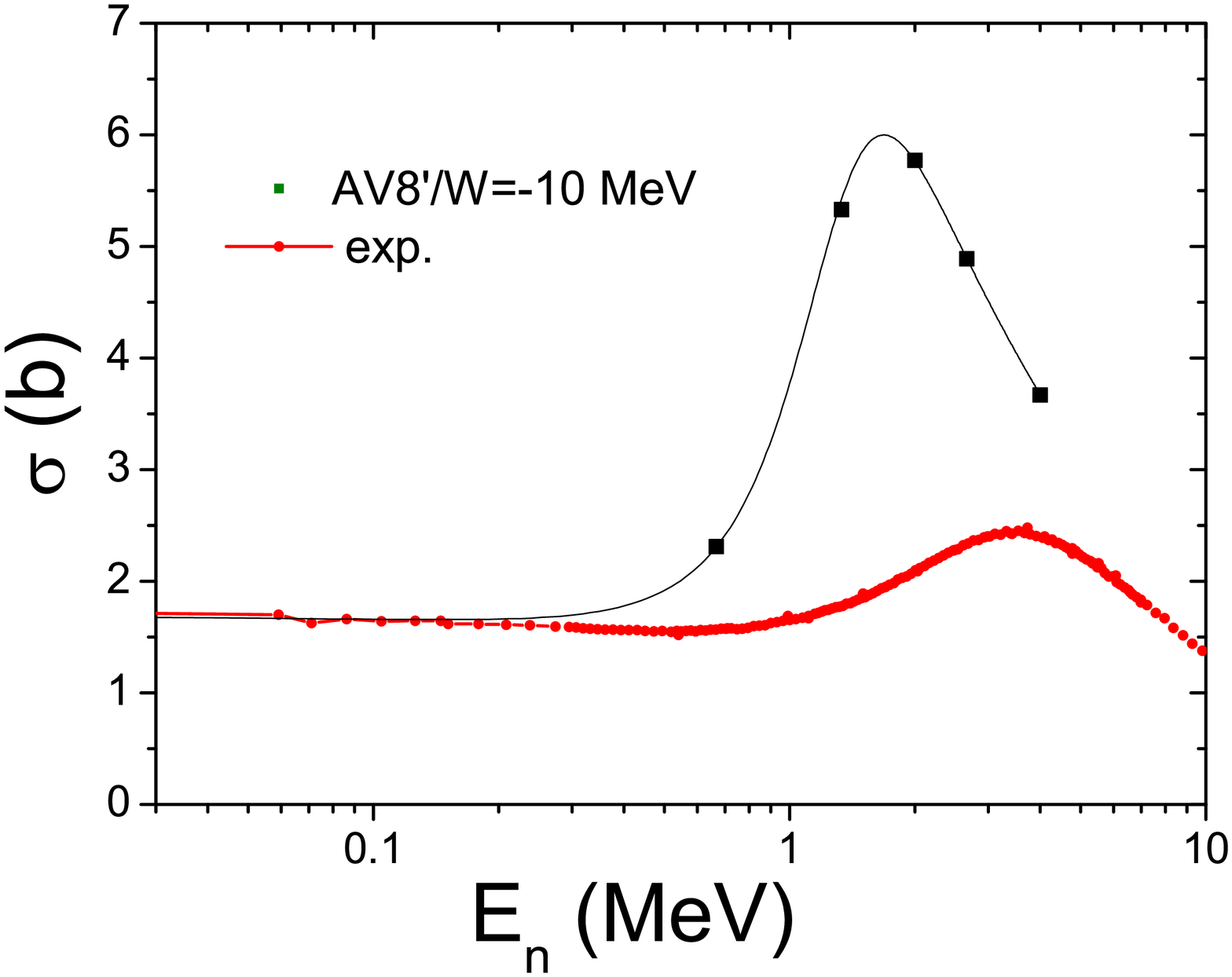,scale=0.30}
\end{center}
\caption{(Color online) The calculated total cross section of $^3$H$+n$
represented by thin black solid line using $W_1(T$=3/2)=$-10$ MeV.
The experimental data~\cite{Phillips-1980} are illustrated by the red thick 
solid line.}
\label{fig:cross-section}
\end{figure}

\vskip 0.1cm 
Results presented in Fig.~\ref{fig:W1-dependence}a,
however, give little insight to the properties of $^4$H, once it
becomes a resonant state for $W_1(T=3/2)>-19$ MeV. Moreover it is
well known~\cite{Tilley1992}, that for broad resonances the
structure given by the S-matrix poles may be different from that
provided by an R-matrix analysis. Therefore, it makes much more sense
to perform direct calculations of the measurable $^3$H$ + n$ data,
namely scattering cross sections. We display in
Fig.~\ref{fig:cross-section} the  $^3$H$ + n$ total cross section
calculated for a value of $W_1(T=3/2)=-10$ MeV. This cross section
is clearly dominated by  pronounced negative-parity resonances in
the $^4$H system. These resonances contribute too much in the
total cross section, resulting in the appearance of a narrow peak shifted
significantly to the lower-energy side. Furthermore, in order to
reproduce the shape of the experimental $^3$H$ + n$ cross section,
a very weak 3NF is required in the isospin $T\!=\!3/2$ channel. From
this fact, we conclude that even a $W_1(T\!=\!3/2)=-10$ MeV value
renders the 3NF to be excessively attractive.

\vskip 0.1cm In conclusion, as far as we can maintain the consistency with
the observed low-lying energy properties of the  $^4$H, $^4$He ($T=1$)
and $^4$Li nuclei, it is difficult to produce an observable $^4n$
resonant state.

\subsection{3n resonances}\label{3n_res}

Finally, in Fig.~\ref{fig:three-neutron}, we show the calculated
resonance poles of the trineutron $^3n$ system for the
lowest-lying negative- and positive-parity states ($J=3/2^-,
1/2^-$ and $1/2^+$). The strength $W_1(T\!=\!3/2)$ is increased so
that there appears a broad  resonance with  Im($E_{\rm res.})
\approx$ $-$Re($E_{\rm res.})$.

\begin{figure}[b]
\vskip 0.5cm
\begin{center}
\epsfig{file=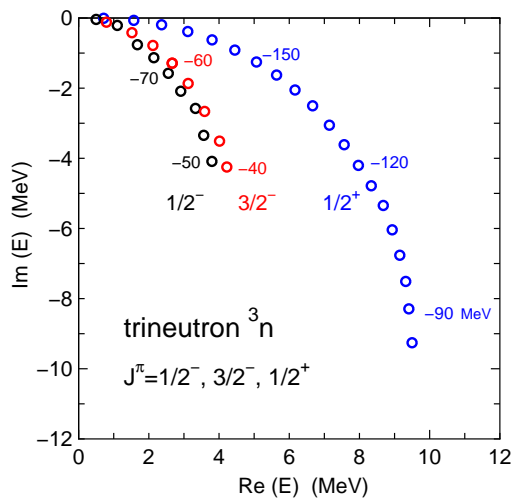,scale=0.40}
\end{center}
\caption{(Color online) Trineutron $^3n$  resonance trajectories for
$J=3/2^-, 1/2^-$ and $1/2^+$ states.
The circles correspond to resonance positions for  $W_1(T=3/2)$
from $-75$ to $-40$ MeV for $J=3/2^-$,
from $-90$ to $-50$ MeV for $J=1/2^-$ and
from $-180$ to $-85$ MeV for $J=1/2^+$ in step of 5 MeV.
}
\label{fig:three-neutron}
\end{figure}

Although it is reasonable to have the negative-parity
states much lower than the positive-parity ones from the viewpoint
of a naive $3n$ shell-model configuration, we have to 
impose a value of $W_1(T=3/2)$, a few
times stronger than in the $^4n$  system to bind  $^3n$.

For $W_1(T\!=\!3/2)=-40$ MeV, the
most favorable $^3n$ resonance with $J=3/2^-$ is located at
Re$(E_{\rm res}) \sim 4$ MeV with $\Gamma \sim 8$ MeV
 as seen in Fig.~\ref{fig:three-neutron}, whereas the  $J=0^+$ $^4n$ state
is still  bound at $E=-4.5$ MeV (cf. Fig.~4b).

On the other hand the $^3n$ is a much more repulsive system than the $^4n$
one, which benefits from the presence of two almost bound bosonic
dineutron pairs. On the other hand $^4n$ is much more sensitive
to the 3NF than $^3n$, as four neutrons involve four 3NF
interactions compared to single one present in three neutrons.
From this result one can expect that if the 3NF contribution turns
out to be important in the $^4n$ system rendering it resonant, other
multineutron systems with $A>4$ (in particular $^6n$ and $^8n$)
would display even more prominent resonant structures than $^4n$.



\section{Conclusions}\label{Conclusions}

Motivated by the recent experimental claim regarding the possible 
existence of  observable tetraneutron
$^4n$~\cite{Kisamori_PhD,Kisamori_PRL} states, we have investigated
the possibility that the $4n$ system exhibits a near-threshold bound 
or narrow resonant state compatible
with the reported data.

\vskip 0.1cm When studying the tetraneutron 
sensitivity to the ingredients of the
nuclear interaction, we have concluded that this system is not
very sensitive to ``experimentally allowed"
modifications in $NN$ interaction. The most natural way to enhance a
tetraneutron system near the threshold is through an additional
attractive isospin $T=3/2$ term in the three-body force. We have
examined the consistency of the nuclear Hamiltonian modifications,
required to produce observable tetraneutron states, with other
four-nucleon observables, like the low-lying $T=1$ states in
$^4$H, $^4$He and $^4$Li.

\vskip 0.1cm This study has been based on the $4N$ Hamiltonian considered 
in Ref.~\cite{Hiya04SECOND}  built from the AV8$'$ version of the
 $NN$ potential produced by the Argonne group and supplemented 
with a phenomenological
$3N$ force including both $T=1/2$ and $T=3/2$ terms. The $T=1/2$
term was adjusted~\cite{Hiya04SECOND} to properly reproduce the
binding energies of the $^3$H and $^3$He ground states and the
$^4$He ground and  first $0^+$ excited states as well as the
transition form factor $^4{\rm He}(e,e')^4{\rm He}(0^+_2)$.
In order to check the model independence of our results we have also considered
the INOY potential  which markedly differs from the preceding ones in its 
non local character that incorporates the $T=1/2$ 3NF.
Despite of these differences the results were very close to each other.

\vskip 0.1cm The  $T=3/2$  component added in the present work, has the same
functional form as the $T=1/2$ one and  contains one single free
parameter -- the strength of its attractive part --  which was
adjusted in order to generate a $4n$ bound or resonant states. The
validity of the strength of $T=3/2$ parameter found in this way,
was investigated by calculating the $T=1$, $J^\pi=2^-$ and $1^-$
states of $^4$H, $^4$He and $^4$Li as well as the total cross
section of the $^3$H$+n$ scattering, which turned out to be very
sensitive to the $T=3/2$ $3N$ force.

The {\it ab initio} scattering solutions of the $4n$ Hamiltonian were obtained using the appropriate boundary condition of a four-body resonance provided by the complex scaling method.
The Gaussian expansion method and the Faddeev-Yakubovsky  formalisms were used  for solving the $A=4$ problem. The two methods  provide
 accurate results and agreed with each other within at least two significant digits, both for the resonant as well as for the bound states.

\vskip 0.1cm
In order to produce resonant tetraneutron states 
which were situated in the complex energy plane close to the physical axis, and, thus may have an observable impact, we were obliged to introduce strong modifications in the 
 $T=3/2$ $3N$ force.
These modifications were however found to be inconsistent with other well
 established nuclear properties and low energy scattering data.
This result is in line with our previous study of the tetraneutron system\cite{Lazauskas-4n}.

In conclusion, we were not able to validate the recent observation of a
$^4n$ signal~\cite{Kisamori_PhD,Kisamori_PRL} as 
related to the existence of resonant  $4n$ states.

Further experimental studies of the tetraneutron system  are planned 
in the near future at RIKEN  \cite{Shimoura,Kisamori,Paschalis}  to  confirm
the finding of~\cite{Kisamori_PhD,Kisamori_PRL} with higher statistic.
If this would be the case it will constitute a real challenge for 
theoretical interpretation.

Another possibility could be that the
low-energy $^4n$ signal is a result of some unknown dynamical
phenomena, which is not directly related to the existence of
$S$-matrix (resonance) poles in the $^4n$ system. In this respect it
is worth mentioning that, in the framework of  the scattering
theory, there exist other possibilities to generate sharp
structures in a reaction cross section without any presence of
$S$-matrix pole singularities~\cite{Calucci}. Such phenomena 
have not yet been established in any physical system.

\section*{Acknowledgments}
The authors thank Dr. S. Shimoura, Dr. K. Kisamori, Dr. H. Sakai,
 Dr. F. M. Marques, Dr. V. Lapoux for valuable discussion.
This work was partially performed in the "Espace de Structure
Nucl\'eaire Th\'eorique" (ESNT, http://esnt.cea.fr)  at CEA  from
which the authors acknowledge support. The numerical calculations
were performed on the HITACHI SR16000 at KEK, YITP in Kyoto
University and  HPC  resources  of TGCC under the allocation
2015-x2015056006 made by GENCI. We thank the staff members of
these computer centers for their constant help.
This work was partly supported by RIKEN iTHES Project .


\end{document}